\documentclass[12pt,a4paper]{article}
\usepackage{graphicx}
\usepackage{bm}
\usepackage{amssymb}
\usepackage{amsmath}
\usepackage{apacite}
\begin{document}

\title{Reinforcement learning of recurrent neural network for temporal coding}

\author{Daichi Kimura \and Yoshinori Hayakawa\thanks{Department of Physics, Tohoku University, Sendai 980-8578, Japan}}
\date{}

\maketitle

\begin{abstract}
We study a reinforcement learning for temporal coding with neural network consisting of stochastic spiking neurons. 
In neural networks, information can be coded by characteristics of the timing of each neuronal firing, 
including the order of firing or the relative phase differences of firing. 
We derive the learning rule for this network and show that the network consisting of Hodgkin-Huxley neurons with the dynamical synaptic kinetics can learn the appropriate timing of each neuronal firing. 
We also investigate the system size dependence of learning efficiency. 
\end{abstract}

\section{Introduction}
Many studies have assumed that neurons transmit information by their firing rate. 
The McCulloch-Pits unit is a typical model and networks of these units have been investigated. 
On the other hand, recent experiments suggest that the timing of neuronal firing may also contribute to the information representation function in the brain and the synaptic modification \cite{gray,bi,varela,reyes}. 
For example, it seems that local and global synchronization play a significant role in integration of information which is distributed across the 
brain. 
Another example shows that the order of timing of neuronal firings can encode the information of stimuli on fingertips, and this encoding by sequence can transmit information faster than coding from the firing rate directly \cite{johansson}. 
    
To capture the dynamical aspects of neural networks, networks consisting of various model neurons other than the McCulloch-Pitts unit have been investigated, 
because McCulloch-Pitts units cannot describe the temporal behavior of neurons over short time scales. 
In this context, an associative memory for neural networks of oscillator neurons or spiking neurons has been studied \cite{aoyagi,yoshioka,hoppensteadt,kanamaru,hasegawa,lee,lee2}. 
In these systems, the relative phase differences, i.e., the timing of firings, are used to represent the memory.

There are few studies of learning in pulse neuron models such as those consisting of Hodgkin-Huxley (HH) neurons 
because of difficulty in deriving the learning rule. 
Although several studies have been made of learning in networks that consist of integrate-and-fire (IF) model neurons \cite{seung,xie,cios}, 
most of these studies focus only on coding in terms of the firing rate. 

However, it would be useful to combine temporal coding and learning because it has been shown that temporal coding can deal with more information and process it faster than coding from just the firing rate \cite{thorpe}. 
As an example of this, Delorme et al.(2001) show that a neural network consisting of IF neurons can learn to identify human faces 
by using ``Rank Order Coding'', i.e., coding by the order of timing of each neuronal firing, 
where neurons are allowed to spike once only. 

In this paper, we study a reinforcement learning for temporal coding with neural network consisting of stochastic spiking neurons.  
After defining a network of coupled stochastic HH neurons and some quantities in Sec. II, 
we train the network to learn an XOR operation, 
where the output information is coded by the order of firing in Sec. III. 
In Sec. IV, we investigate how the result or performance of learning depends on the system size and the strength of noise, 
and conclusions follow.  
 
\section{The model}

\begin{figure}[tbp]
  \begin{center}
    \includegraphics[keepaspectratio=true,height=80mm]{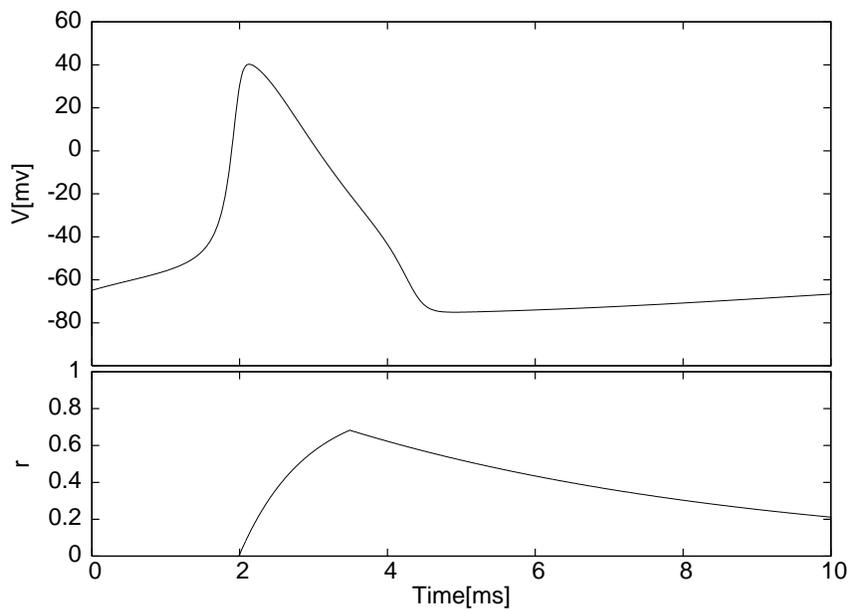}
  \end{center}
  		\caption{The behavior of $V_i(t)$ and $r_j(t)$ of single HH neuron added the external current. The neuron fires at $t=2$ ms, then the value of $r_i(t)$ starts to increase. After a lapse of $\tau=1.5$ ms, $r_i(t)$ turns into decline.}
  \label{fig:hh.eps}
\end{figure}

To illustrate an example of the learning process of spiking neurons, 
we consider a neural network consisting of HH neurons.
Since HH neurons show excitability, they can code information by the timing of firing. 
The complete dynamics for a network of coupled HH neurons may be expressed as
\begin{gather}
C_{m} \frac{dV_{i}}{dt}=g_{Na}m_{i}^{3}h_{i}(V_{Na}-V_{i})+g_{K}n_{i}^{4}(V_{K}-V_{i})+g_{L}(V_{L}-V_{i}) \nonumber\\
+\sum_{j}w_{ij}I_{j}^{s}(t)+\xi_{i}(t), \label{eq:hhnetv} \\
\frac{dm_{i}}{dt}=\alpha_{m}(V_{i})(1-m_{i})-\beta_{m}(V_{i})m_{i}, \label{eq:hhnetm}\\
\frac{dh_{i}}{dt}=\alpha_{h}(V_{i})(1-h_{i})-\beta_{h}(V_{i})h_{i}, \label{eq:hhneth}\\
\frac{dn_{i}}{dt}=\alpha_{n}(V_{i})(1-n_{i})-\beta_{n}(V_{i})n_{i}, \label{eq:hhnetn}
\end{gather}
where $V_{i}$ is the membrane potential of neuron $i$, $C_{m}$ the membrane capacitance, $V_{r}(r=\textrm{Na,K,L})$ are the equilibrium potentials, 
$g_{r}(r=\textrm{Na,K,L})$ the conductance, $m_i,h_i,n_i$ the voltage dependent activating or inactivating variables, 
$\alpha_x$ and $\beta_x(x=m,h,n)$ the functions of voltage $V_{i}$ \cite{hodgkin},
$w_{ij}$ is the synaptic weight from neuron $j$ to $i$ ($w_{ij} \neq w_{ji}$ in general), $I_{j}^{s}(t)$ the synaptic current and 
$\xi_i(t)$ is the Gaussian white noise which obeys,
\begin{gather}
\overline{\xi_{i}(t)}=0, \label{eq:gauss1}\\
\overline{\xi_{i}(t)\xi_{j}(t')}=Q\delta_{ij}\delta(t-t'), \label{eq:gauss2}
\end{gather}
where $\overline{A}$ is the average of $A$ over time and  
$Q$ the variance of noise.
The synaptic current $I_{j}^{s}(t)$ is given by
\begin{equation}
I_{j}^{s}(t)=r_j(t)[V_{syn}-V_{i}], \label{eq:synapticI}
\end{equation}
where $V_{syn}$ is the synaptic reversal potential and $r_j(t)$ the fraction of bound receptors \cite{destexhe} described by
\begin{gather}
\frac{dr_j}{dt}=\alpha T(t)(1-r_j)-\beta r_j, \\
T(t)=\begin{cases}
1 & t_j^0\le t<t_j^0+\tau,\\
0 & \textrm{otherwise},
\end{cases}
\end{gather}
where $\alpha=0.94\ \textrm{ms}^{-1}$, $\beta=0.18\ \textrm{ms}^{-1}$, 
$t_j^0$ the time when the presynaptic neuron $j$ fires (membrane potential over $27$ mV) and $\tau =1.5$ ms \cite{lago}
Fig.~\ref{fig:hh.eps} shows the behavior of $V_i(t)$ and $r_i(t)$ of single neuron added the external current whose amplitude is $10$ mA.   
We used the forth order Runge-Kutta method with the time step $\Delta t=0.01$ to solve Eqs.~(\ref{eq:hhnetv})$\sim$(\ref{eq:hhnetn}).

To train the neural network, we use a reinforcement learning algorithm. 
Let us consider time sequences of states of neurons; 
$\sigma \equiv (\bm{V}(0),\bm{V}(1),\bm{V}(2),\cdots,\bm{V}(T))$, 
where $\bm{V}(t)$ denotes the vector $(V_{1}(t),\cdots,V_{N}(t))$. 
We assign a scalar value (``reward'') to each time sequence $\sigma$ according to the signal from the network \cite{sutton}. 
We give a high reward $R$ to the desirable time sequence $\sigma$ in each episode. 
Here we consider episodic learning. 
Since Eq.~(\ref{eq:hhnetv}) includes the Gaussian white noise, we calculate the expected value of the reward $\langle R\rangle$, 
where $\langle \cdots \rangle$ signifies the average over all possible time sequences $\sigma$. 
Then the goal of learning is to maximize $\langle R\rangle$ by adjusting $w_{ij}$. 
We use an ascending gradient strategy; 
\begin{gather}
w_{ij}^{\mathrm{New}}=w_{ij}^{\mathrm{Old}}+\delta w_{ij}, \label{eq:wupdate} \\
\delta w_{ij}=\epsilon \frac{\partial \langle R\rangle}{\partial w_{ij}}, \label{eq:koubaikouka}
\end{gather}
where $\epsilon$ is the learning coefficient.
We can calculate the gradient of $\langle R\rangle$ with respect to $w_{ij}$ \cite{hayakawa,fiete},
\begin{equation} 
\frac{\partial \langle R\rangle}{\partial w_{ij}}=\frac{1}{Q}\left\langle R(\sigma)\int_{0}^{T}dt\xi_{i}(t)I_{j}^{s}(t)\right\rangle .\label{eq:gakusyufhn}
\end{equation}
For details of the derivation, see appendix.

\section{Learning procedure for temporal coding}

The present learning rule Eq.~(\ref{eq:wupdate})$\sim$(\ref{eq:gakusyufhn}) can be effective for any information coding including the order coding. 
To show an example, we consider a neural network consisting of $2$ input neurons, $15$ output neurons and hidden neurons. 
We divide the set of output neurons into three disjoint subsets, 
$\mathcal{O}_{1}$, $\mathcal{O}_{2}$ and $\mathcal{O}_{3}$, each containing 5 output neurons. 

Here, we assume that information is coded by the temporal order of the ``group activity'' of subsets $\mathcal{O}_{1}$, $\mathcal{O}_{2}$ and $\mathcal{O}_{3}$. 
As a learning goal, we chose an XOR operation in terms of order coding; 
the subsets $\mathcal{O}_{1}$, $\mathcal{O}_{2}$ and $\mathcal{O}_{3}$ should fire in this order for input patterns $[-1,-1]$ and $[1,1]$, and in reverse order for input patterns $[-1,1]$ and $[1,-1]$. 
We train the neural network to learn the suitable timing $t_{p}^{k}$ of firing in the subset $\mathcal{O}_{p}$ for given input pattern $k=[-1,-1],[-1,1],[1,-1]$ and $[1,1]$, respectively, as described in Table \ref{tab:order}. 
\begin{table}[tbp]
 \caption{The order of collective firing as a learning goal. $t_{1}=2.0$, $t_{2}=4.0$ and $t_{3}=6.0$, respectively.}
 \begin{center}
  \begin{tabular}{|c|c|c|c||c|}
    \hline
    Input &\multicolumn{3}{|c||}{$t_{p}^{k}$} &Desirable \\ \cline{2-4}
    pattern &$\mathcal{O}_{1}$ &$\mathcal{O}_{2}$ &$\mathcal{O}_{3}$ &order\\
    \hline
    $[-1,-1]$ &$t_1$ &$t_2$ &$t_3$ &$\mathcal{O}_{1}$, $\mathcal{O}_{2}$, $\mathcal{O}_{3}$    \\
    
    $[1,-1]$ &$t_3$ &$t_2$ &$t_1$  &$\mathcal{O}_{3}$, $\mathcal{O}_{2}$, $\mathcal{O}_{1}$    \\
    
    $[-1,1]$ &$t_3$ &$t_2$ &$t_1$  &$\mathcal{O}_{3}$, $\mathcal{O}_{2}$, $\mathcal{O}_{1}$    \\
    
    $[1,1]$ &$t_1$ &$t_2$ &$t_3$  &$\mathcal{O}_{1}$, $\mathcal{O}_{2}$, $\mathcal{O}_{3}$    \\
    \hline
  \end{tabular}
 \end{center}
 \label{tab:order}
\end{table}
The right column of Table \ref{tab:order} shows the desirable order of group activity of subsets for each input pattern.
The learning goal is the reduction of the RMS errors $E^{k}$ for each input pattern $k$ given by
\begin{equation}
E^{k}=\frac{1}{N_{\textrm{O}}}\sum_{p=1}^{3}\sum_{i\in {O}_{p}}\left(t_{i}^{0}-t_{p}^{k}  \right)^{2},  
\end{equation}
where $t_i^0$ is the time when the neuron $i$ fires and $N_{\textrm{O}}$ the number of output neurons.

We define an ``episode'' as the transient dynamics of the network which lasts from $t=0$ to $t=T$ accompanied with external inputs. 
We evaluate the reward $R$ depending on the resulting error $E^{k}$ through each episode for given input pattern as 
\begin{equation}
R=\begin{cases}
1 & E^{k}<E_{\textrm{th}}^{k},\\
0 & E^{k}\ge E_{\textrm{th}}^{k},
\end{cases} \label{eq:reward}
\end{equation}
where $E^{k}_{\textrm{th}}$ is a threshold of error. 
We change the threshold $E^{k}_{\textrm{th}}$ gradually as the learning proceeds.
In the following simulations, we calculate the average error $E_{\textrm{ave}}^{k}$ over $100$ episodes for each input pattern, 
and we update the threshold for the next $100$ episodes by $E_{\textrm{th}}^{k}=0.99\times E_{\textrm{ave}}^{k}$. 
$E_{\textrm{th}}^{k}$ and $E_{\textrm{ave}}^{k}$ are almost identical. 
It sometimes happens that $E_{\textrm{ave}}^{k}$ increases through learning since the update of $w_{ij}$ to decrease $E^k$ will affect the value of $E^{k'}$ where $k\neq k'$. 
Thus, $E_{\textrm{th}}^{k}$ also sometimes increases, not decreases monotonically.

During each episode, the input signals are fixed to the time-independent value, 
$I_{i}^{s}=I_{0}$ or $-I_{0}$, corresponding to the input patterns. (Here, $I_{0}=10$ mA.) 
The learning coefficient, the variance of noise and the period of one episode are $\epsilon =0.001$, $Q=10000$ and $T=10.0$ ms, respectively.

We assume that all the output and hidden neurons are excitatory, 
i.e., $w_{ij}$ for $j\not\in \mathcal{I}$ can take only positive value, where $\mathcal{I}$ is the set of input neurons, 
while $w_{ij}$ for $j\in \mathcal{I}$ can take both positive and negative value.  
We first set the synaptic weights $w_{ij}$ for $j\in \mathcal{I}$ using the uniformly distributed random number in $[-1.0,1.0]$, 
while for $j\not\in \mathcal{I}$ between $[\frac{2.5}{N},\frac{5.0}{N}]$, where $N$ is the number of neurons. 
Due to this normalization factor $\frac{1}{N}$, the magnitude of error $E^k$ does not depend on $N$. 
In this section, the number of hidden neurons was $5$.
 
\begin{figure}[tbp]
  \begin{center}
    \begin{tabular}{c}
      \resizebox{!}{80mm}{\includegraphics{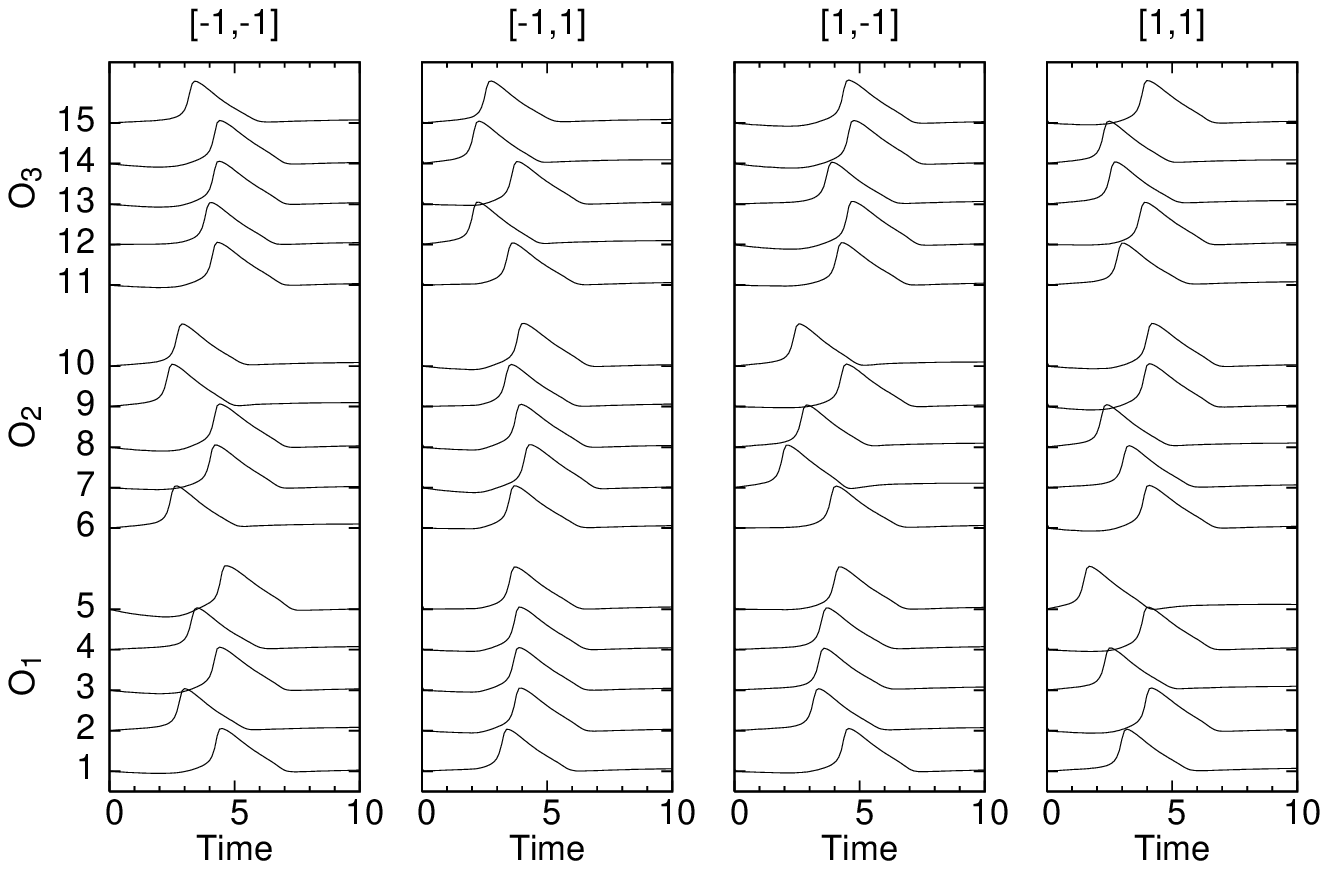}} \\
      \resizebox{!}{80mm}{\includegraphics{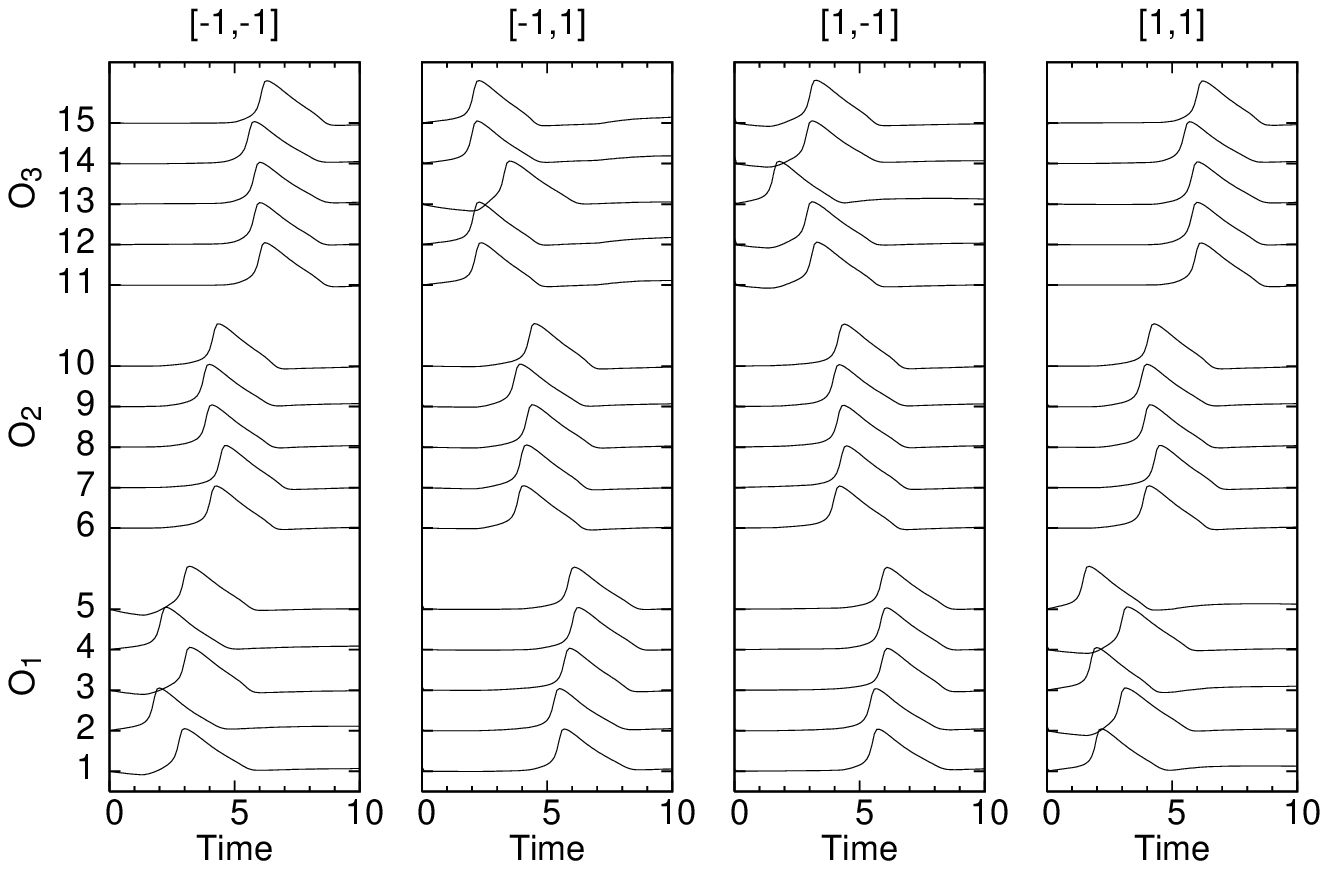}} \\
    \end{tabular}
	\caption{The action potential of each output neuron at initial state (upper) and after $5\times 10^4$ episodes for each input pattern (that is, $2\times 10^5$ episodes in total) elapsed (lower). The index of each output neuron is shown on the vertical axis. The neurons with index $\le 5$ are in $\mathcal{O}_{1}$, those with index $\ge 11$ are in $\mathcal{O}_{3}$, and the others are in $\mathcal{O}_{2}$. $\mathcal{O}_{1}$, $\mathcal{O}_{2}$ and $\mathcal{O}_{3}$ fire in this order for input patterns [-1,-1] and [1,1], while in reverse order for [1,-1] and [-1,1]. See also Table \ref{tab:order}.}
	\label{fig:xor.eps}
  \end{center}
\end{figure}

\begin{figure}[htbp]
  \begin{center}
  \includegraphics[keepaspectratio=true,height=80mm]{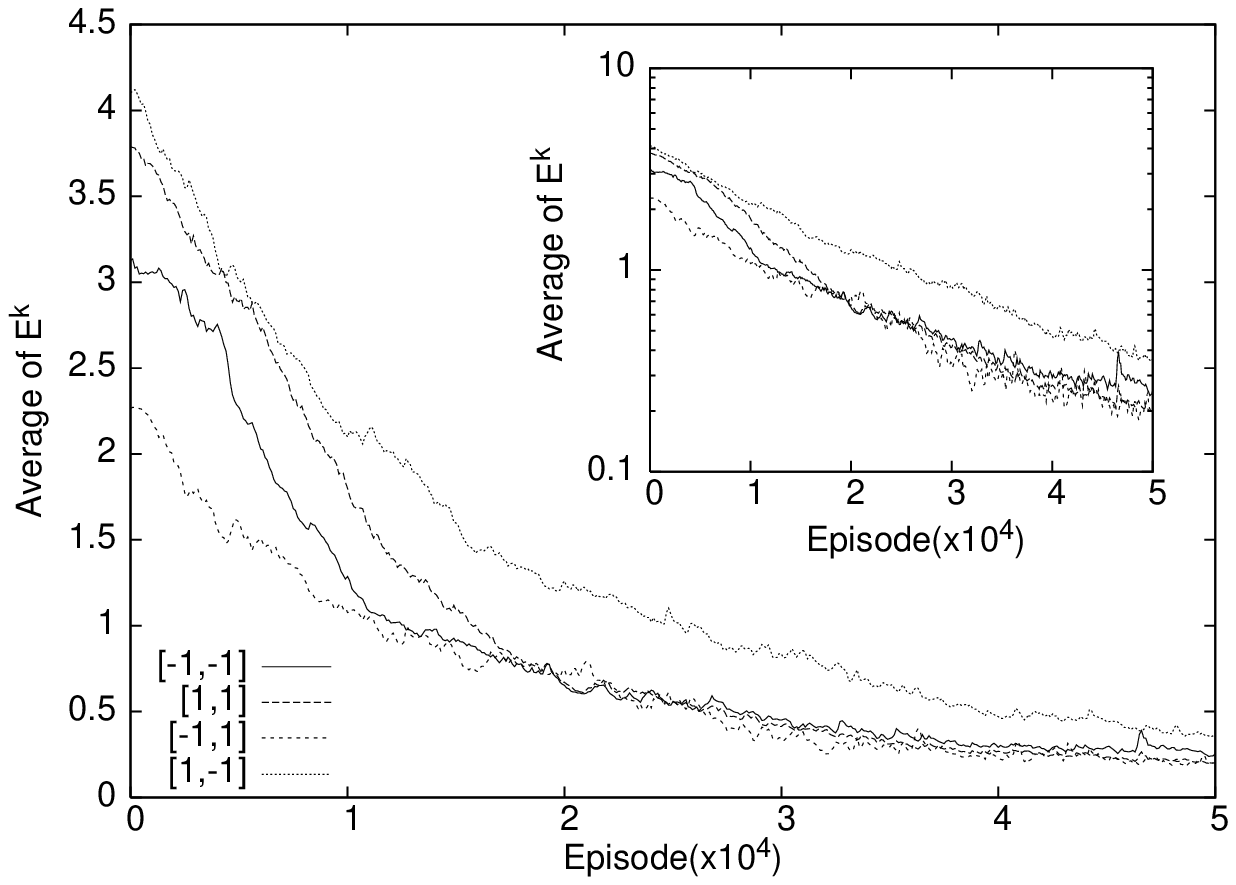}
  \end{center}
  \caption{$E_{\textrm{ave}}^k$ for each input pattern as a function of episodes through learning. Inset is the same data in semi-logarithmic plot. Typical value of $E^{k}$ before learning is approximately $3$.}
  \label{fig:xorE.eps}
\end{figure}

Fig.~\ref{fig:xor.eps} shows the behavior of the output neurons at initial state (upper) and 
that after $5\times 10^4$ episodes for each input pattern (that is, $2\times 10^5$ episodes in total) elapsed (lower). 
We find that neurons in $\mathcal{O}_{1}$, $\mathcal{O}_{2}$ and $\mathcal{O}_{3}$ fire in this order for the input patterns $[-1,-1]$ and $[1,1]$, 
while in reverse order for the patterns $[1,-1]$ and $[-1,1]$ after learning. 
In fact, Fig.~\ref{fig:xorE.eps} shows that the average error $E_{\textrm{ave}}^k$ decrease gradually through learning. 
These figures indicate that learning in terms of the order of firing by applying the present learning rule is successful. 

\begin{figure}[tbp]
  \begin{center}
    \includegraphics[keepaspectratio=true,height=70mm]{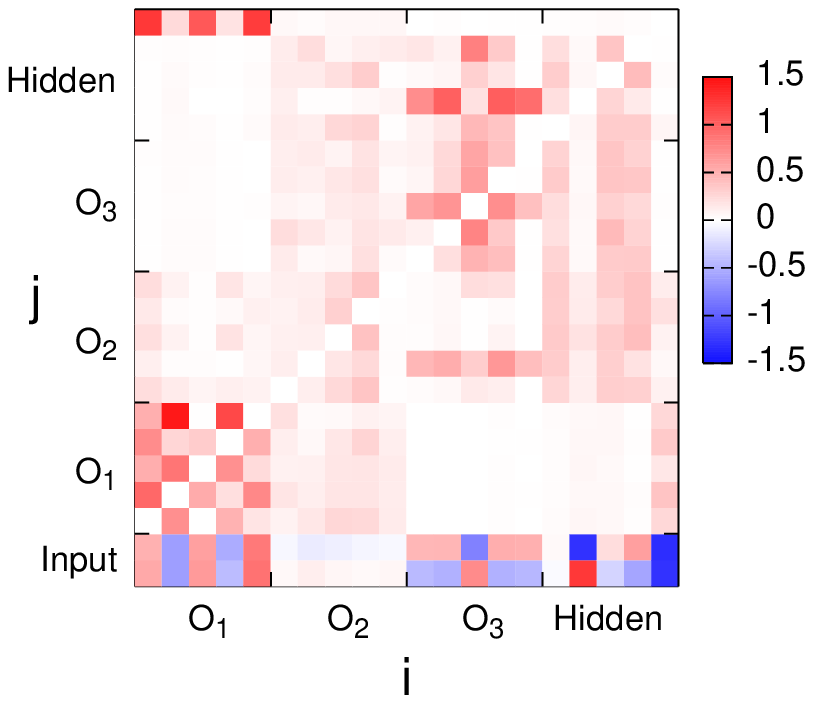}
    \caption{$w_{ij}$ in a matrix form representing the connections with color after learning. Neuron $j$ is presynaptic and $i$ is postsynaptic.}
    \label{fig:matrix.eps}
  \end{center}
\end{figure}

Fig.~\ref{fig:matrix.eps} shows $w_{ij}$ in a matrix form representing the connections with color after learning. 
As can be seen, there is few connection between $\mathcal{O}_{1}$ and $\mathcal{O}_{3}$ as a result of learning. 
This can be interpreted as the consequence that there is no direct correlation between $\mathcal{O}_{1}$ and $\mathcal{O}_{3}$ 
in the both of the output time sequence.  
Few connections from $\mathcal{I}$ to $\mathcal{O}_{2}$ can be understood in the same way.  

\section{System size dependence of learning efficiency}

Since realistic biological systems consist of many neurons, it is interesting and important how efficient this learning rule is for larger system size. 
In this section, we discuss the system size dependence of the performance of learning

\begin{figure}[tbp]
  \begin{center}
    \begin{tabular}{c}
      \resizebox{!}{80mm}{\includegraphics{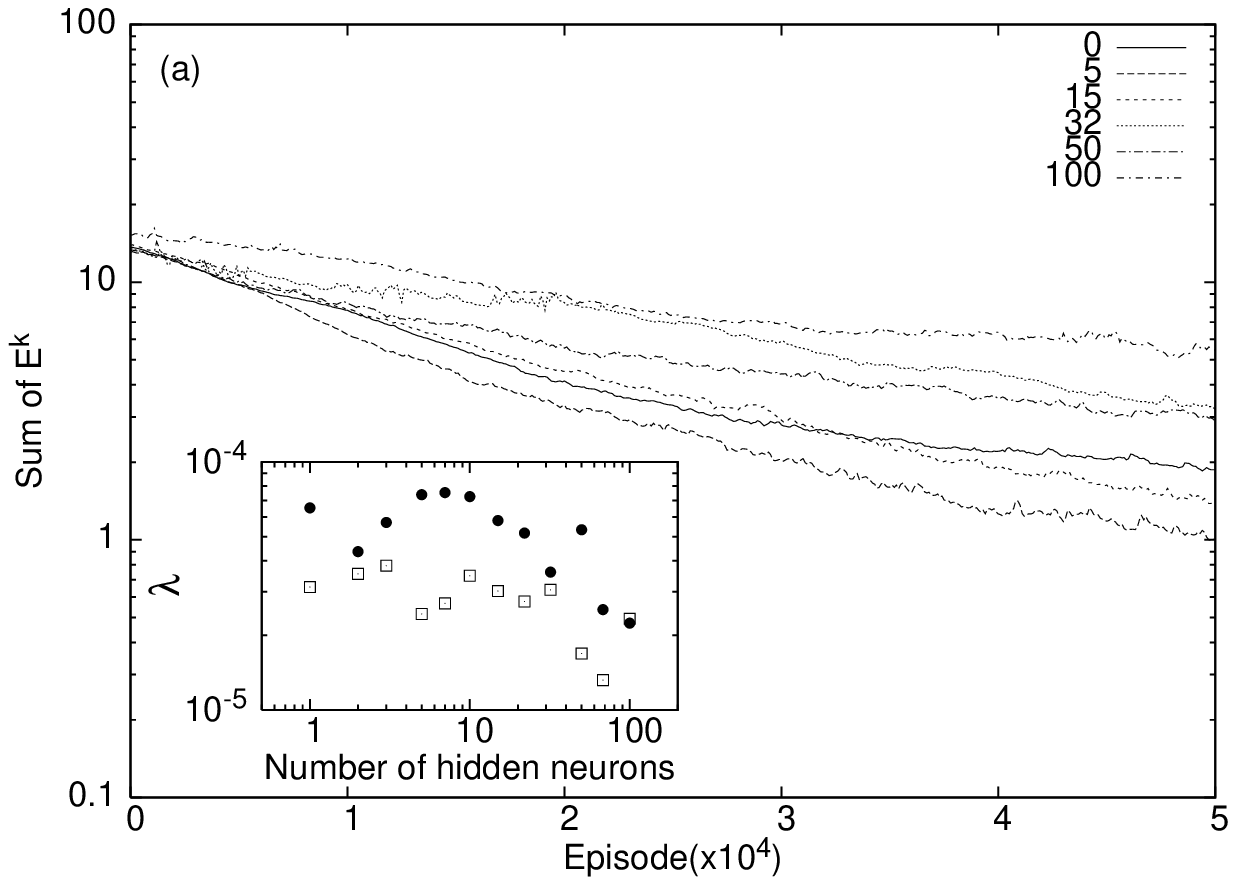}} \\
      \resizebox{!}{80mm}{\includegraphics{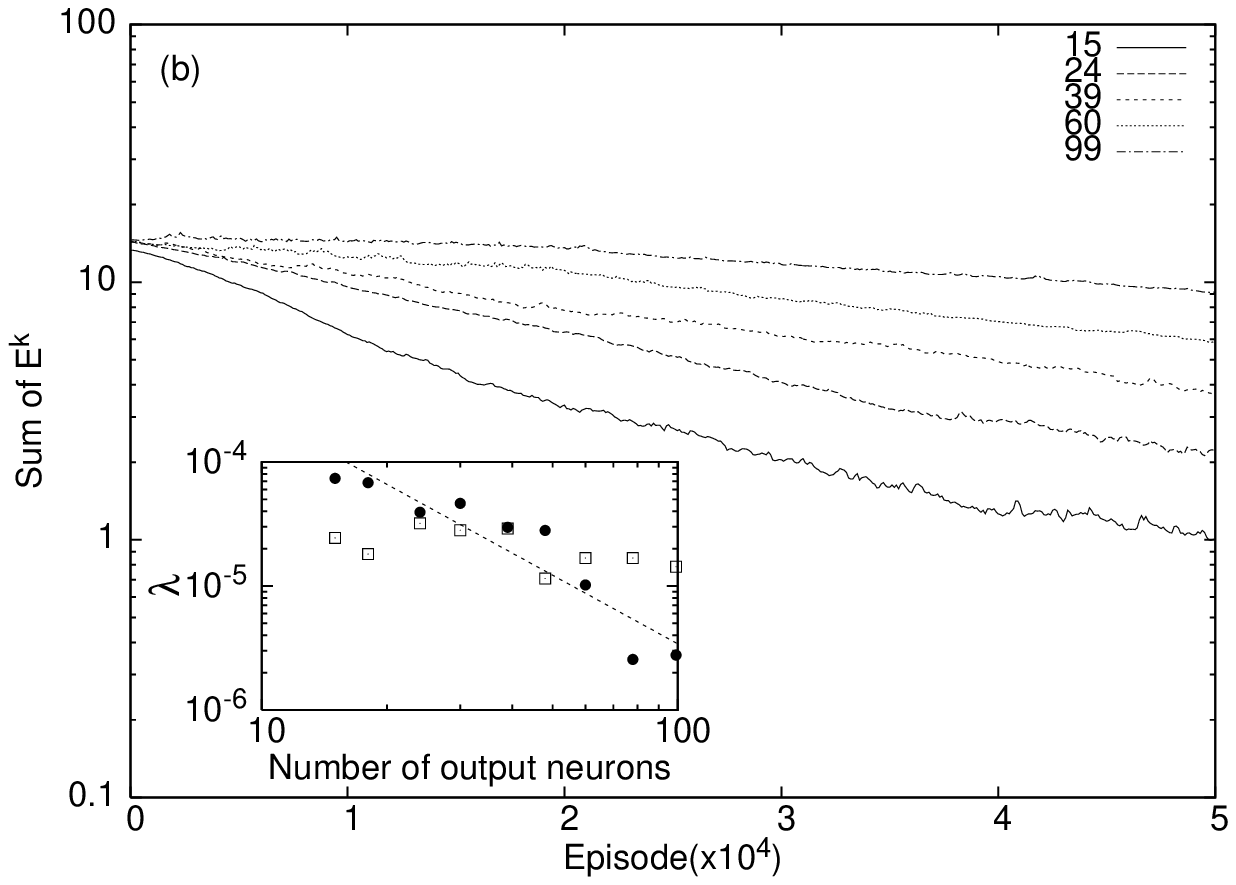}} \\
    \end{tabular}
	\caption{$\sum_{k}E_{\textrm{ave}}^k$ as a function of episodes through learning for each $N_{\textrm{H}}$(a) and $N_{\textrm{O}}$(b). Inset: The decay constant $\lambda$ of $\sum_{k}E_{\textrm{ave}}^k$ for $0<s<10^4$ (filled circle) and $4\times 10^4<s<5\times 10^4$ (open square) where $s$ is the number of episode. The slope of dotted line in inset (b) is approximately $-2$.}
	\label{fig:errorsize.eps}
  \end{center}
\end{figure}

At first, we train the network for various number of hidden neurons $N_{\textrm{H}}$. 
In the following simulation, all parameters but $N_{\textrm{H}}$ are given as the same as previous section. 
The sum of average errors $\sum_{k}E_{\textrm{ave}}^k$ are shown in Fig.~\ref{fig:errorsize.eps}(a).  
To investigate the property of learning, we assumed an exponential decay of the error $\sim \exp(-\lambda s)$ as the function of the number of episode $s$   and showed $\lambda$ in inset of Fig.~\ref{fig:errorsize.eps}(a). 
As shown in Fig.~\ref{fig:errorsize.eps}(a), 
the speed of learning becomes slower when extra hidden neurons are added, and, decay rate $\lambda$ also has the same tendency. 
However, the error is significantly reduced as the episodes proceed. 
In this case, it turned out that the network with $N_{\textrm{H}}=5$ showed best result in this simulation 
so that those added neurons are redundant and do not contribute learnability.

Next, we examine the case the learning speed for various number of output neurons $N_{\textrm{O}}$, where each subsets $\mathcal{O}_{p}$ consist of $N_{\textrm{O}}/3$ and the number of hidden neurons $N_{\textrm{H}}=5$. 
Fig.~\ref{fig:errorsize.eps}(b) shows the error $\sum_{k}E_{\textrm{ave}}^k$ and the decay rate $\lambda$ (inset). 
In this first stage of the learning process ($s\le 10^4$), decay rate is approximately given as $\lambda \sim N_{\textrm{O}}^{\delta}$, 
where $\delta\approx -2$.
However, if $\lambda$ is evaluated in the interval $4\times 10^4<s<5\times 10^4$, 
the $N_{\textrm{O}}$ dependence of $\lambda$ seems to be more weaken, 
so that a long-term behavior of the learning process needs to be investigated to discuss the asymptotics.  

\begin{figure}[tbp]
  \begin{center}
    \includegraphics[keepaspectratio=true,height=80mm]{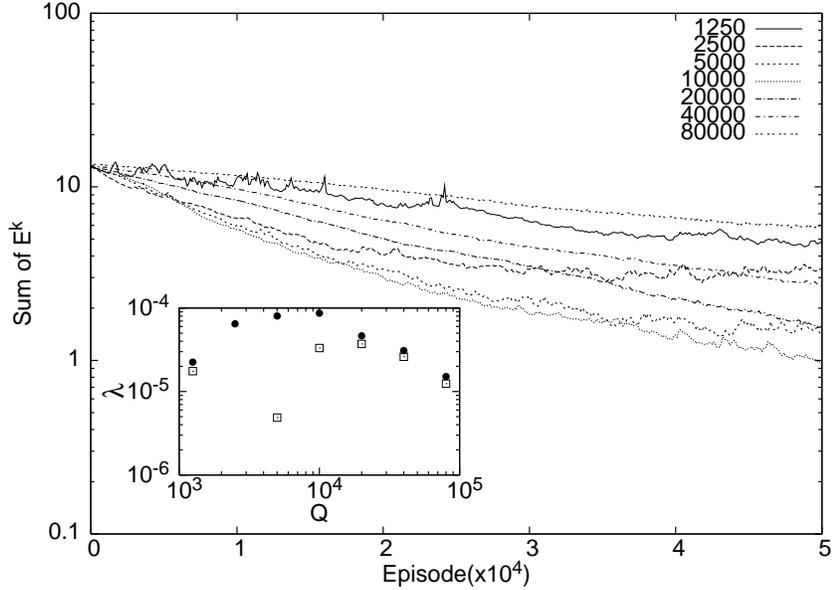}
  \end{center}
  \caption{$\sum_{k}E_{\textrm{ave}}^k$ as a function of episodes through learning for each $Q$. Inset: $\lambda$ of $\sum_{k}E_{\textrm{ave}}^k$ for $0<s<10^4$ (filled circle) and $4\times 10^4<s<5\times 10^4$ (open square).}
  \label{fig:errornoise.eps}
\end{figure}

We would like to mention the effects of noise in the learning. 
The noise strength may affect the result or performance of learning 
because a random search process in the weight space is the essential part of this learning algorithm. 
We tested the dependence of the learning performance on the variance of noise $Q$ in Eq.~(\ref{eq:gauss2}) , 
where $N_{\textrm{O}}=15$, $N_{\textrm{H}}=5$ and other parameters are given as previous section. 
Fig.~\ref{fig:errornoise.eps} shows the learning error and its decay rate. 
Since the amount of update $|\delta w_{ij}|$ is in the order of $|\epsilon/Q|$, 
the magnitude of $Q$ will directly affect the stability of learning. 
However, Fig.~\ref{fig:errornoise.eps} indicates this learning rule is robust against the change of $Q$ 
as long as the gradient ascending process assumed in Eq.~(\ref{eq:koubaikouka}) holds.

\section{Conclusions}

We have proposed a concrete method to encode temporal information for a network of stochastic spiking neurons utilizing a reinforcement learning method, 
and we show that this learning rule is effective for order coding. 

Since the derivation of Eq.~(\ref{eq:koubaikouka}) does not depend on the form of the differential equation for the action potential 
like Eq.~(\ref{eq:hhnetv}), 
the learning rule Eq.~(\ref{eq:koubaikouka}) is generally applicable to networks of other types of model neurons, 
for example neurons following the IF or FitzHugh-Nagumo (FHN)  model with random fluctuation. 
In fact, we have checked the network consisting of IF or FHN neurons can also learn the same order coding task with the learning rule Eq.~(\ref{eq:koubaikouka}). 
(We omit these details in this paper.)

The present learning rule can be applied not only to the order coding, but also to any information coding. 
In general, if one can design an appropriate reward $R$ by defining the error of group activity from the reference signals, 
the learning process is given in a straightforward manner as described in Sec. III. 
For example,
we confirmed that the present method also worked for the XOR task in terms of phase coding, 
where the relative phase difference codes information in the same way as oscillator neurons \cite{aoyagi}. 
In the simulation, neural network consisted of $2$ input neurons, a few hidden neurons and $10$ output neurons  
which were divided into two disjoint subsets, $\mathcal{O}_{1}$ and $\mathcal{O}_{2}$, each containing 5 output elements. 
Learning goal was for group activity in $\mathcal{O}_{1}$ and $\mathcal{O}_{2}$ to become in phase for input patterns $[-1,-1]$ and $[1,1]$, 
and to be out of phase for $[1,-1]$ and $[-1,1]$. 

In general, convergence of the reinforcement learning is slow and the present model is not the exception. 
Several factors need to be investigated to improve the speed of convergence.
Although we have assumed Gaussian white noise in this study, 
in real neuronal systems it is likely that noise may obey other statistical properties. 
For example, Fiete et al. (2006) discusses a learning rule using the arbitrary noise.  
Furthermore, learning speed depends to a great extent on the design of the cost function, i.e., the reward.
If the design of the reward is not suitable, the learning process drops into local minima easily. 
In the present study, we empirically employed a sliding threshold method by changing the learning goal $E_{\textrm{th}}^{k}$ in Eq.~(\ref{eq:reward}) 
started from easier one. 
Although this strategy can be applied to any learning tasks, 
the performance or result of learning is sensitive to define $E_{\textrm{th}}^{k}$. 
In this paper, we have defined $E_{\textrm{th}}^{k}$ as $A\times E_{\textrm{ave}}^{k}$ where $A=0.99$, because $A=0.9$ is too severe to learn so that the speed of learning becomes slower and $A=1.0$ causes overestimated reward so that the learning process drops into local minima. 
The point that $E_{\textrm{th}}^{k}$ does not monotonically decrease is also important 
since the update of $w_{ij}$ to decrease $E^k$ will affect the value of $E^{k'}$ where $k\neq k'$. 
Therefore, if we decrease $E_{\textrm{th}}^{k}$ monotonically while $E_{\textrm{ave}}^{k}$ increases because of this influence with update of $w_{ij}$, 
the task becomes harder to learn. 
As well as the conventional back propagation learning, 
how to decide the number of hidden neurons and assign the initial value of $w_{ij}$ are also important problem. 
In the XOR task with the order coding, convergence speed in learning process significantly affected by both the number of hidden and output neurons, 
larger size system seems to require more learning steps. 
We intend to examine this relationship in future research.

We have assumed a fully connected network and assigned the random value to the initial state of $w_{ij}$, 
which may not be appropriate for learning. 
Watts and Strogatz have proposed a ^^ ^^ small-world'' network structure \cite{watts}. 
This network has been investigated in the context of a multi-layered feed-forward network, and some improvements in performance have been made \cite{simard}. 
Also, the propagation velocity and coherent oscillation of IF or Hodgkin-Huxley neurons depend on the topology of network 
(which includes small-world network) \cite{lago,roxin}.
These results may indicate that there are some suitable topologies of network for temporal coding as well. 
Since the present learning rule Eq.~(\ref{eq:gakusyufhn}) includes only the local relation between neuron $i$ and $j$, 
this learning rule may be applied to a network with an arbitrary topology. 
Finding a suitable network topology to facilitate learning is also a problem for future investigation.

\section*{Acknowledgments}
We would like to thank Dr. Toshihiro Kawakatsu and Dr. Tsuyoshi Hondou for helpful comments.

\section*{Appendix : Estimation of gradient of $\langle R\rangle$}
Let us introduce a stochastic processing element as a model neuron.
The value of action potential of $i$-th neuron at discrete time $t (=0,1,2,\cdots)$ is represented by $V_{i}(t)$.
$V_{i}(t)$ and other internal state $m_{i}(t)$ at successive time step $t+1$ are then determined as
\begin{gather}
V_{i}(t+1)=g(V_{i}(t),m_{i}(t))+\sum_{j}^{N}w_{ij}f(V_{j}(t))+\xi_{i}(t), \label{eq:discrete} \\
m_{i}(t+1)=h(V_{i}(t),m_{i}(t)), \label{eq:discretev}
\end{gather}
where $w_{ij}$ is the synaptic weight from neuron $j$ to $i$,
$f(V_{i}(t))$ the signal which $i$-th neuron emits according to $V_{i}(t)$,
$N$ the total number of neurons,  
$g(V_{i}(t),m_{i}(t))$ and $h(V_{i}(t),m_{i}(t))$ are functions of the neuronal states,  
and $\xi_{i}(t)$ is the Gaussian white noise which obeys Eq.~(\ref{eq:gauss1}) and Eq.~(\ref{eq:gauss2}). 
We consider that $f(V_{i}(t))$, $g(V_{i}(t),m_{i}(t))$ and $h(V_{i}(t),m_{i}(t))$ are arbitrary functions. 

As a continuous time representation, 
we employ a Langevin equation for neuronal updates;
\begin{gather}
\frac{dV_{i}}{dt}=g(V_{i}(t),m_{i}(t))+\sum_{j}^{N}w_{ij}f(V_{j}(t))+\xi_{i}(t), \label{eq:continuous} \\
\frac{dm_{i}}{dt}=h(V_{i}(t),m_{i}(t)). \label{eq:continuousv}
\end{gather}
Although we treat $m_{i}(t)$ as a scalar variable in the following discussion, 
the same method can also be applied if $m_{i}(t)$ is multivariate. 

We consider time sequences of states of neurons for $t=0,\cdots,T$; \\
$\sigma \equiv (\bm{V}(0),\bm{V}((1),\bm{V}(2),\cdots,\bm{V}(T))$, 
where $\bm{V}(t)$ denotes the vector \\ $(V_{1}(t),\cdots,V_{N}(t),m_{1}(t),\cdots,m_{N}(t))$. 

We assign a reward $R$ to each time sequence $\sigma$ and calculate the gradient of $\langle R\rangle$ with respect to $w_{ij}$ \cite{hayakawa,fiete}. 
The probability density $\mathcal{T}_{\sigma}$ for each time sequence is defined as
\begin{equation}
\mathcal{T}_{\sigma}=P(\bm{V}(T)|\bm{V}(T-1))\cdots P(\bm{V}(1)|\bm{V}(0)), \label{eq:historyP}
\end{equation}
where $P(A|B)$ is the transition probability from state $B$ to $A$. 
Since $m_{i}(t)$ is deterministic and only $V_{i}(t)$ includes the Gaussian noise described by Eqs.~(\ref{eq:gauss1}) and (\ref{eq:gauss2}), 
the transition probability in Eq.~(\ref{eq:historyP}) may be expressed as 
\begin{gather}
P(\bm{V}(t+1)|\bm{V}(t))=\prod_{i}\frac{1}{(2\pi Q)^{1/2}}\exp\left[-\frac{1}{2Q}\xi_{i}^{2}(t)\right] \nonumber \\
\times \delta_{m_{i}(t+1),h(V_{i}(t),m_{i}(t))} \nonumber \\
=\frac{1}{(2\pi Q)^{N/2}}\exp\left[-\frac{1}{2Q}\sum_{i}\eta_{i}(t)\right]\prod_{i}\delta_{m_{i}(t+1),h(V_{i}(t),m_{i}(t))},
\end{gather}
where
\begin{equation}
\eta_{i}(t)\equiv \left[V_{i}(t+1)-g(V_{i}(t),m_{i}(t))-\sum_{j}^{N}w_{ij}f(m_{j}(t))\right]^{2}. 
\end{equation}
and $\delta_{m,n}$ is the Kronecker delta.  
The average of a quantity $Y(\sigma)$ over all possible time sequences can be described by 
\begin{gather}
\langle Y \rangle = \frac{1}{Z}\sum_{\sigma}Y(\sigma)\mathcal{T}_{\sigma} \nonumber \\
=\frac{1}{Z}\sum_{\sigma}Y(\sigma)\exp\left[-\frac{1}{2Q}\sum_{t}\sum_{i}\eta_{i}(t)\right], \label{eq:expectedY}
\end{gather}
where $Z$ is a normalization factor. 
Therefore, the gradient of $Y(\sigma)$ may be obtained as
\begin{gather}
\frac{\partial \langle Y\rangle}{\partial w_{ij}}=\frac{1}{Z}\sum_{\sigma}\mathcal{T}_{\sigma}Y(\sigma)\frac{1}{Q}\sum_{t=0}^{T}\xi_{i}(t)f(V_{j}(t)) \nonumber \\
=\frac{1}{Q}\left\langle Y(\sigma)\sum_{t=0}^{T}\xi_{i}(t)f(V_{j}(t))\right\rangle .
\end{gather}

In the continuous case (\ref{eq:continuous}), the sum over $\sigma$ in Eq.~(\ref{eq:expectedY}) becomes the path integral 
\begin{equation}
\langle Y \rangle =\frac{1}{Z}\int_{\sigma}Y(\sigma)\exp \left[-\frac{1}{2Q}\int_{0}^{T}dt\sum_{i}\tilde{\eta_{i}}(t)\right]d(\textrm{path}), \label{eq:pathintegral}
\end{equation}
where
\begin{equation}
\tilde{\eta_{i}}(t)\equiv \left[\frac{dV_{i}}{dt}-g(V_{i}(t),m_{i}(t))-\sum_{j}^{N}w_{ij}f(V_{j}(t))\right]^{2}. 
\end{equation}
Using $R(\sigma)$ as $Y(\sigma)$, 
we obtain the derivative in Eq.~(\ref{eq:koubaikouka}) as  
\begin{equation} 
\frac{\partial \langle R\rangle}{\partial w_{ij}}=\frac{1}{Q}\left\langle R(\sigma)\int_{0}^{T}dt\xi_{i}(t)f(V_{j}(t))\right\rangle .\label{eq:gakusyu}
\end{equation}
For the network consisting of HH neurons (\ref{eq:hhnetv})$\sim $(\ref{eq:hhnetn}), 
we can derive Eq.~(\ref{eq:gakusyufhn}) by instituting Eq.~(\ref{eq:synapticI}) to Eq.~(\ref{eq:gakusyu}).

Thus, this result indicates that the calculation of the gradient of $\langle R\rangle$ depends only on  $\xi_i(t)$, $f(V_j(t))$ and $R(\sigma)$. 
The linear summation of the pre-synaptic neuronal activity and an additive independent noise as in Eq.~(\ref{eq:discrete}) and Eq.~(\ref{eq:continuous}) is essential for the present learning algorithm. 
This learning rule includes only the local relation between neurons $i$ and $j$, 
while back-propagation \cite{rumelhart} or RTRL \cite{williams} requires the information of other neurons in addition to $i,j$ in order to calculate $\delta w_{ij}$.

\nocite{delorme}
\bibliographystyle{apacite}
\bibliography{FHN}

\end{document}